\documentclass{emulateapj}
\usepackage{graphicx}
\newcommand{\tmin}{{T_{\rm min}}}

\begin{document}
\title{The Journey Counts:  The Importance of Including Orbits when Simulating Ram Pressure Stripping}
\author{Stephanie Tonnesen$^{1}$, }
\affil{$^{1}$Flatiron Institute, CCA, 162 5th Avenue, New York, NY 10010 USA}
\email{1 stonnes@gmail.com (ST)} 

\abstract{We investigate the importance of varying the ram pressure to more realistically mimic the infall of a cluster satellite galaxy when comparing ram pressure stripping simulations to observations. We examine the gas disk and tail properties of stripped cluster galaxies in eight ``wind-tunnel" hydrodynamical simulations with either varying or constant ram pressure strength.   In simulations without radiative cooling, applying a varying wind leads to significantly different density and velocity structure in the tail than found when applying a constant wind, although the stripping rate, disk mass, and disk radius remain consistent in both scenarios.  In simulations with radiative cooling, the differences between a constant and varying wind are even more pronounced.  Not only is there a difference in morphology and velocity structure in the tails, but a varying wind leads to a much lower stripping rate, even after the varying wind has reached the ram pressure strength of the constant wind.  Also, galaxies in constant and varying wind simulations with the same gas disk mass do not have in the same gas disk radius.  A constant wind cannot appropriately model the ram pressure stripping of a galaxy entering a cluster.  We conclude that simulations attempting detailed comparisons with observations must take the variation of the ram pressure profile due to a galaxy's orbit into consideration.}

}
\section{Introduction}

Cluster satellite galaxies may undergo a number of interactions that are specific to dense environments. These include interactions between the intracluster medium (ICM) and the galaxy, such as ram pressure stripping and starvation (Gunn \& Gott 1972; Larson, Tinsley \& Caldwell 1980), and gravitational interactions such as those between galaxies, like harassment (Moore et al. 1996), and between a galaxy and the cluster potential, such as tidal stripping (Merritt 1984; Gnedin 2003). The relative influence of these mechanisms in transforming galaxy morphology, color, and gas content remains unclear, with large surveys helping to disentangle the various drivers (e.g. Moran et al 2007; van den Bosch et al. 2008). An alternative way to gauge their relative importance, and to get a deeper understanding of the processes themselves, is to look for observational signs of the different interactions in individual galaxies.

There are a few signatures that can be used to differentiate between gas-stripping mechanisms like ram pressure stripping and gravitational interactions.  For example, late-type galaxies in the center of the Virgo cluster have smaller H I disks than stellar disks, indicating an interaction that does not affect the stellar component of galaxies (Cayatte et al. 1990; Warmels 1988; but see Smith et al. 2012 for simulations showing gas dragging the stellar disk). Studies of H I deficiency have shown that galaxies in clusters have less neutral hydrogen than their counterparts in the field (see the review by Haynes et al. 1984).  Ram pressure stripped galaxies should therefore have small gas disks and correspondingly lower star formation rates (Gavazzi et al. 2006; Koopmann \& Kenney 2004).  However, ram pressure stripping may also increase star formation rates in the surviving gas disk (Dressler \& Gunn 1983; Evrard 1991; Fujita 1998; Smith et al. 2010; Poggianti et al. 2004; 2009; 2017; Fujita \& Nagashima 1999; Tonnesen \& Bryan 2012; Bekki 2014; Kapferer et al. 2009). 

Perhaps the strongest indicator of ram pressure stripping is single-sided gas tails, as observed in HI, H$\alpha$, X-ray, and molecular emission (e.g.   Irwin \& Sarazin 1996; Oosterloo \& van Gorkom 2005; Haynes et al. 2007; Chung et al. 2007, 2009; Kenney et al. 2008; Yoshida et al. 2008; Yagi et al. 2010; Sun et al. 2006; Jachym et al. 2014, 2017; Boselli et al. 2018; Lee \& Chung 2018; George et al. 2018; Moretti et al. 2018a; Moretti et al. 2018b; Poggianti et al. 2019).  Recently, Poggianti et al. (2016) found ram pressure stripping candidates by identifying unilateral disturbances in optical emission.  

With the advent of integral field units, detailed observational maps of galaxies have become possible.  For example,  Merluzzi et al. (2013; 2016) used the integral field spectrograph WiFeS and imaging data to map the kinematics and physical conditions of the ionized gas and stellar populations of galaxies with signatures of ram pressure stripping in the Shapley supercluster.  Using MUSE (Multi Unit Spectroscopic Explorer) spectroscopy, other researchers have mapped other cluster galaxies (e.g. Fossati et al. 2016; Poggianti et al. 2017; Bellhouse et al. 2017; Gullieuszik et al. 2017; Moretti et al. 2018a).

In addition, numerical simulations can be used to predict observational signatures of stripped galaxies.  For example, Bekki (2014) ran several simulations, varying galaxy and cluster mass as well as galaxy orbits and inclinations, to study how ram pressure stripping generally affects star formation rates and H$\alpha$ emission.  This work concluded that star formation rates and H$\alpha$ distributions in the galaxy disk are affected by ram pressure in a variety of ways depending on the galaxy mass, inclination angle, and ram pressure strength. 

On the other hand, simulations have also been used to determine if observed galaxies have been ram pressure stripped.  This has been done extensively for Virgo cluster galaxies using an N-body code (Vollmer et al. 2001; 2003; 2006; 2008; 2009; 2012).  For example, in Vollmer et al. (2008), the authors model 4 ram pressure profiles, each with 4 different disk-wind inclination angles to find a best match to NGC 4501.  More recently, Merluzzi et al. (2013) focus on a galaxy in the Shapley supercluster, and use N-body/hydrodynamical simulations to verify their proposed ram pressure stripping scenario.  They run a grid of more than 100 simulations, varying the inclination angle between the galaxy and the ICM wind, the wind velocity, and the gas disk scale height (see also Merluzzi et al. 2016).  

In this paper we consider one aspect of simulating ram pressure stripping in galaxies:  varying ram pressure over time to mimic the infall of a satellite galaxy.   The importance of varying ram pressure has been studied with regards to gas removal in elliptical galaxies for decades.  Takeda et al. (1984) find that as a galaxy falls into the cluster gas stripping can be quite efficient, removing much of the galaxy's gas from the outside-in in the form of a smooth blob of stripped gas.  Toniazzo \& Schindler (2001) model a range of orbits and find that stripping is most efficient when ram pressure increases strongly to a high value, and otherwise gas stripping proceeds more slowly.  Recently Roediger et al. (2015a,b) have examined the stripping of elliptical galaxies in detail using M89 as a reference point, and find that the details of the remaining gas and tail properties depend on the galaxy potential, initial gas distribution, galaxy orbit and orbital stage as well as ICM plasma properties. 

Significant work has also been done studying the importance of varying ram pressure on stripping of spiral galaxies.  In addition to work specifically modeling individual galaxies (e.g. Vollmer et al. 2001; 2003; 2006; 2008; 2009; 2012), Jachym et al. (2007, 2009) find that the total amount of ICM sweeping past stripped galaxies is more important than the peak ram pressure encountered, and that while more highly inclined disks tend to have less gas removed this difference is eliminated in strongly stripped galaxies that encounter a large ICM column density.  Roediger \& Bruggen (2007) find that in an orbiting galaxy, gas is lost more slowly than in an instantaneous prediction, although the remaining radius and total gas mass stripped is similar to the Gunn \& Gott (1972) analytic estimate as long as the inclination is not high.  Roediger \& Bruggen (2008) find that the tail mass distribution depends on the galaxy orbit.  

However, because a galaxy's orbit is uncertain, it is often ignored when comparing observations to simulations (e.g. Tonnesen et al. 2011; Merluzzi et al. 2013; 2016; Gullieuszik et al. 2017).  Particularly if a galaxy is still falling into a cluster, and so has only experienced increasing ram pressure, this simplification may be based on the assumption that because ram pressure stripping is a fast process, only the peak ram pressure a galaxy experiences determines the amount of gas stripped.  In this work we focus on whether constant ram pressure stripping simulations can be used to model observed orbiting galaxies by directly comparing simulations with a constant ram pressure to those with a varying ram pressure, focusing on the first infall of a galaxy, so only increasing the ram pressure.  We find that a constant ram pressure cannot simultaneously reproduce both the disk and tail properties produced by a varying ram pressure profile.  Therefore, we argue that simulators must include varying ram pressure due to galaxy infall in order to directly compare with observed galaxies.

The organization of this paper is as follows.  In Section \ref{sec:method} we describe our simulation method, with Sections \ref{sec:galaxy} and \ref{sec:sims} detailing the galaxy model and the individual simulation parameters, respectively.  We then examine the results of our simulations, first focusing on the stripping rate in Section \ref{sec:disk_results}.  In the following results sections we focus on properties of the disk (Sec \ref{sec:gasradius}) and tail (Sec \ref{sec:tail_results}) that can be compared to observations.  We compare simulations with and without radiative cooling to understand the physics behind our results in Section \ref{sec:discussion}.  Finally, we summarize our conclusions in Section \ref{sec:conclusions}.

\section{Methodology}\label{sec:method}

We use the adaptive mesh refinement (AMR) code {\it Enzo} (Bryan et al. 2014).   To follow the gas, we employ an adaptive mesh for solving the fluid equations including gravity.  The code begins with a fixed set of static grids and automatically adds refined grids as required in order to resolve important features in the flow.

Our simulated region is 160 kpc on a side with a root grid resolution of $256^3$ cells.   In the central 80 kpc we allow an additional 4 levels of refinement, for a smallest cell size of 39 pc.  We refine the grid based on the local gas density, and choose parameters that refine most of the galactic disk to 39 pc resolution.  

Simulations including radiative cooling use the Sarazin \& White (1987) cooling curve, with no star formation or heating processes.  To mimic effects that we do not model directly (such as stellar and supernovae feedback, sub-grid turbulence, UV heating, magnetic field support, or cosmic rays), we cut off the cooling curve at a minimum temperature $T_{\rm min}$ so that the cooling rate is zero below this temperature.  In these simulations we use $\tmin = 8000$ K. 

To analyze our data we use yt, a toolkit for analyzing and visualizing quantitative data (Turk et al. 2011).  We use yt to create projections and slices, as well as to select disk gas both spatially and using gas density and/or a passive tracer.  yt is then able to perform analysis tasks on the selected data. 

\subsection{The Galaxy}\label{sec:galaxy}

Our galaxy is placed at the center of our computational volume, and remains stationary throughout the runs.  The lower x, y, and z boundaries are all set to inflow in the ICM wind runs, and the wind direction varies depending on the run.  The upper x, y, and z boundaries are set to outflow.

We model a massive spiral galaxy with a flat rotation curve of 205 km s$^{-1}$.  It consists of a gas disk that is followed using the adaptive mesh refinement algorithm (including self-gravity of the gas), as well as the static potentials of the stellar disk, stellar bulge, and dark matter halo.  We follow Roediger \& Br\"uggen (2006) in our modeling of the stellar and dark matter potential and gas disk.  Specifically, we model the stellar disk as a Plummer-Kuzmin disk (Miyamoto \& Nagai 1975), the stellar bulge as a spherical Hernquist profile (Hernquist 1993), and the dark matter halo as the spherical Burkert (1995) model (see Mori \& Burkert 2000 for the analytic potential).  We describe our disk model in detail in Tonnesen \& Bryan (2009, 2010).   In this paper our stellar disk has a radial scale length of 3.5 kpc, a vertical scale length of 0.7 kpc and a total mass of 1.15$\times$10$^{11}$ M$_{\odot}$; the stellar bulge has a scale length of 0.6 kpc and a total mass of 10$^{10}$ M$_{\odot}$; and the dark matter halo has a scale radius of 23 kpc and a central density of $3.8 \times 10^{-25}$ g cm$^{-3}$.  The gas disk has a mass of 8$\times$10$^{10}$ M$_{\odot}$, and radial and vertical scale lengths of 7 kpc and 0.9 kpc, respectively.

To identify gas that originated in the galaxy we follow a passive tracer that is initially set to 1.0 inside the galaxy and $10^{-10}$ outside.  

\subsection{The Simulations}\label{sec:sims}

\begin{table}
\scriptsize
\caption{}\label{table-runs}
\begin{tabular}{cccccc}\hline

Run ID & Rad. & Face-on & Wind & Initial & Max.\\
 & Cooling? & Wind? & Profile? & Press. & RP\\
\hline
RCVW & yes & no & vary & 9.84e-14 & 1.337e-11\\
RCFOVW & yes & yes & vary & 9.84e-14 & 1.337e-11\\
RCCW & yes & no & const & 2.79e-12 & 1.337e-11\\
RCFOCW & yes & yes & const & 2.79e-12 & 1.337e-11\\
RCFOCWL & yes & yes & const& 2.09e-12 & 1.001e-11\\
RCFOCWD & yes & yes & const & 2.79e-12 & 1.337e-11\\
NCFOVW & no & yes & vary & 9.84e-14 & 1.337e-11\\
NCFOCW & no & yes & const & 2.79e-12 & 1.337e-11\\

\end{tabular}
\\
\\
Details of the simulations discussed in this paper.  All units are cgs.
\end{table}

In this paper we discuss eight simulations, summarized in Table \ref{table-runs}.  All of the simulations initially have the same galaxy density profiles, and allow the galaxy to evolve in a static surrounding medium.  

\begin{figure}
\begin{center}
\includegraphics[scale=0.5]{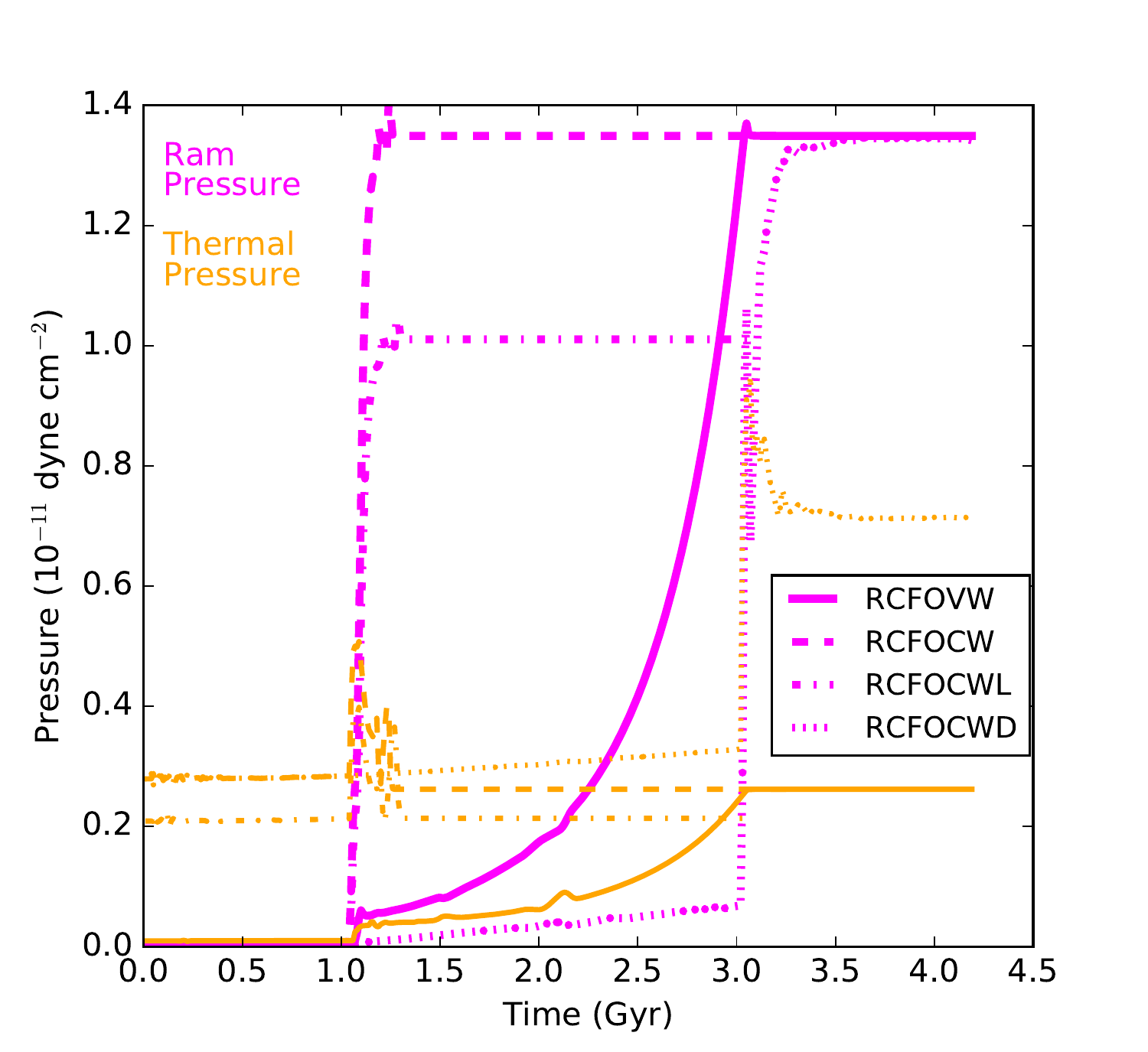}
\caption{Ram pressure and thermal pressure as a function of time near the simulation box edge at the midplane of the disk.  The dashed lines are the constant wind runs (CW) and the solid line is the varying wind run (VW).  Although we only show the measured values for the radiatively cooled cased with a face-on wind (``RCFO"), the runs with wind coming in at an angle and without radiative cooling show nearly identical profiles.  The ram pressure is in thicker magenta lines and the thermal pressure is in thinner orange lines.  Note that the ram pressure is measured as the density multiplied by the total velocity squared, which means that before the wind hits the galaxy there can be ``ram pressure" measured from gas cooling onto the galaxy.  We clearly see that the thermal pressure is less than the peak ram pressure across all runs.  The ram pressure reaches its peak value after about 1.2 Gyr in the CW runs and 3.1 Gyr in the VW runs. }\label{fig-winds}
\end{center}
\end{figure}

In seven of our simulations, after 1 Gyr we generate an ICM inflow.   The one simulation with a delayed wind is denoted with a ``D".  For the six runs that include radiative cooling, denoted by ``RC" in the Run ID (see Table \ref{table-runs}), this time allows cool, dense gas to form in the galaxy ($\rho$ $\ge$ 10$^{-22}$ g cm$^{-3}$).  This naturally generates a multiphase ISM (see Tasker \& Bryan (2006) and Tonnesen \& Bryan (2009) for more discussion of the ISM properties).    

Six of the simulations use a wind that is moving along the rotation axis of the galaxy (denoted by ``FO"), while two simulations model a wind at a 53$^{\circ}$ angle.  Three of the simulations have a wind that increases in strength (denoted by ``VW").  Four simulations have a constant wind:  three at the maximum ram pressure of the varying winds, and one at 75\% of the maximum varying ram pressure (denoted by ``CW" or ``CWL").  In Figure \ref{fig-winds} we show the different wind ram pressure strength profiles for comparison.  Note that our varying wind profiles (``VW") are stripped for an extra Gyr at the maximum ram pressure for comparison purposes.

The ICM conditions are selected such that the initial wind has a Mach number of about two so that the initial wind hitting the galaxy is well described by the shock-jump conditions.  Together with our ram pressure profiles, the Mach number sets our initial thermal ICM pressure as denoted in Table \ref{table-runs}.  Although the initial thermal ICM pressures differ by more than an order of magnitude because the initial ram pressure in the ``VW" runs is lower, the thermal pressure is always much lower than the peak ram pressure experienced by the galaxies (see also Figure \ref{fig-winds}).

To briefly summarize our nomenclature, all runs are identified as ``RC" or ``NC", indicating radiative cooling or no cooling.  Also, all runs either have ``VW" or ``CW" indicating varying wind or constant wind.  Finally, many of our simulation names include ``FO" indicating that the wind is face-on.

Although our results are general and do not depend on the details of the varying wind, we briefly explain how we derive the ram pressure profile here.  We model a galaxy orbiting a cluster using galpy, a python package for orbital dynamics (Bovy 2015, http://github.com/jobovy/galpy).  We model the cluster as an NFW potential with a virial mass of 4.41e14 M$_{\odot}$, and virial radius of 1.55 Mpc.  We use a concentration of 4, as this is a reasonable fit to most clusters (Mandelbaum et al. 2008).  We assume that the cluster is spherically symmetric and static in order to simplify the model.  We can then create a series of possible orbits, and choose one for this work.  The ICM density is modeled as a beta-profile.  The peak ram pressure that we model is found when the galaxy velocity is 1500 km/s as it infalls 1.4 Mpc from the cluster center, so this particular galaxy orbit would have increasing ram pressure as the galaxy continued towards pericenter passage.  The wind begins 2 Gyr before this point, at a distance of 2.9 Mpc from the cluster center (within 2 virial radii of this cluster). 

\begin{figure}
\begin{center}
\includegraphics[scale=0.5]{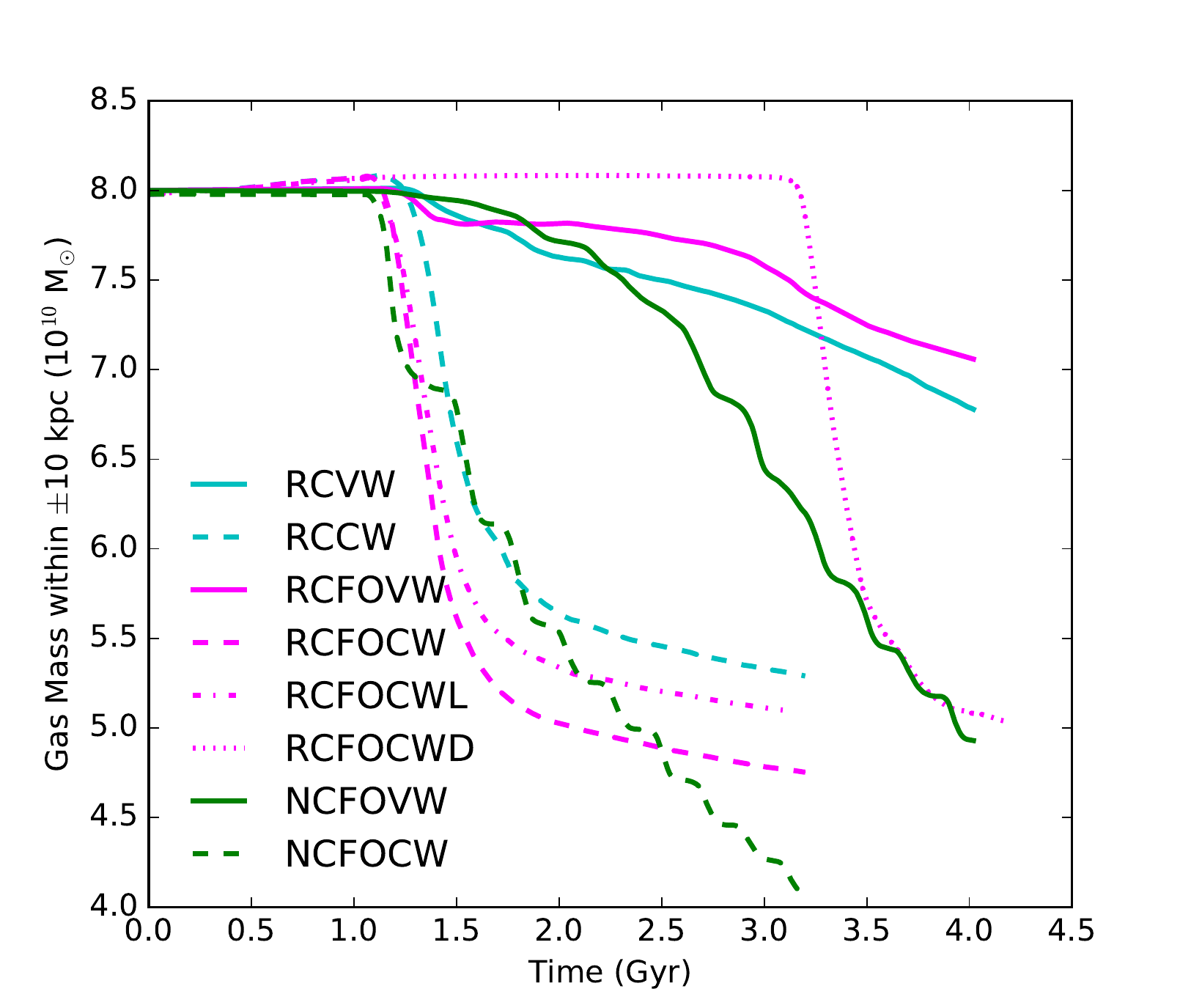}
\caption{The gas mass as a function of time for all seven runs performed for this work.  As in Figure \ref{fig-winds}, the solid and broken lines denote the VW and CW runs, respectively.  Line color denotes run type as in the legend.  Note the dramatic difference in the gas mass between the VW and CW runs, particularly for the simulations that include radiative cooling. }\label{fig-diskmass}
\end{center}
\end{figure}

\section{Results}\label{sec:results}
\subsection{Gas Disk}\label{sec:disk_results}

The importance of ram pressure stripping as a galaxy quenching mechanism strongly depends on the amount of gas it can remove from the disk.  Therefore, we first consider the amount of gas removed by these different wind profiles.  In Figure \ref{fig-diskmass} we plot the disk gas mass, defined as gas with a tracer fraction of more than 0.5 within $\pm$ 10 kpc of the disk plane, as a function of time.  

\begin{figure}
\includegraphics[scale=0.33]{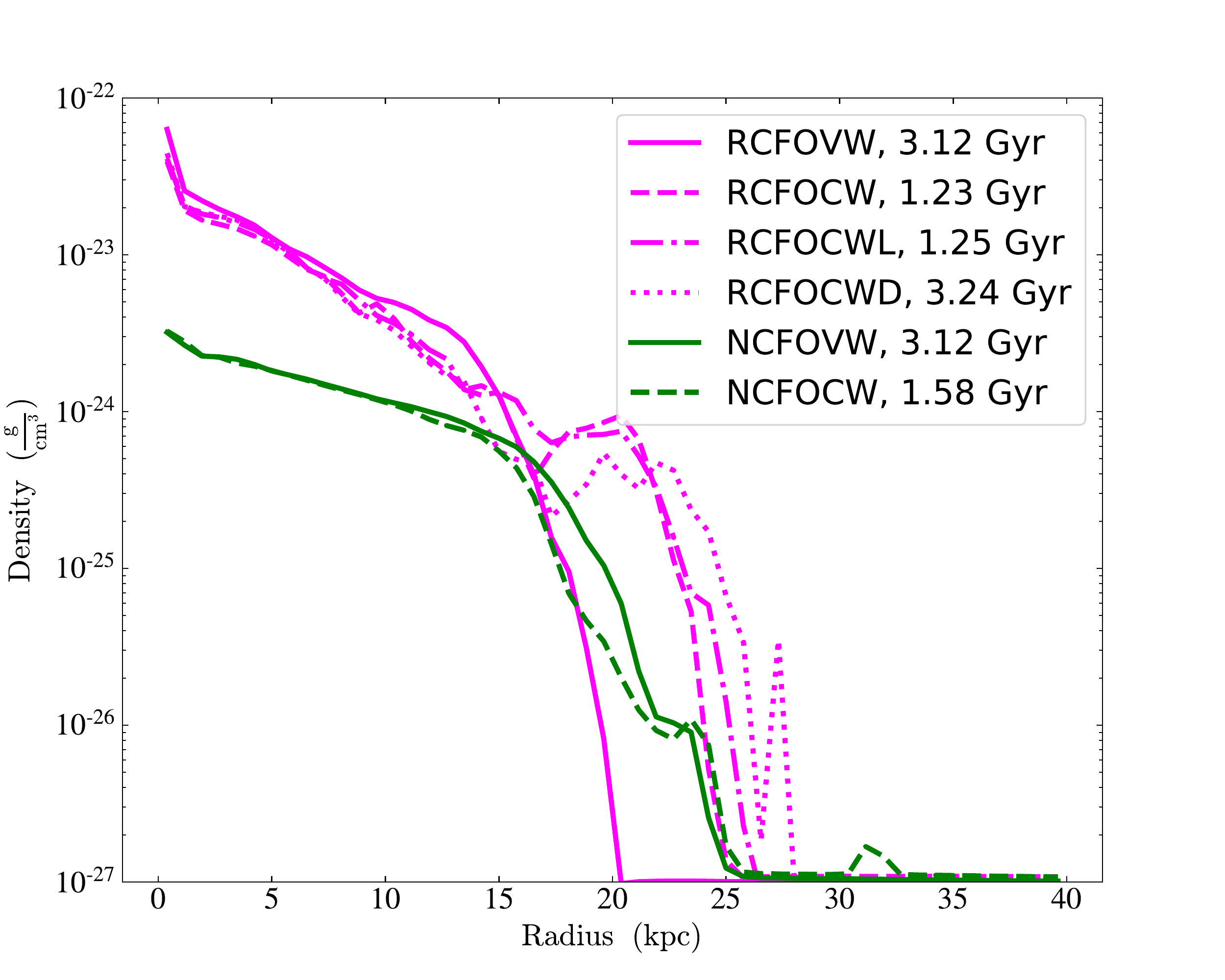}
\caption{Average gas density profiles within the central $\pm$2 kpc of the disk plane.  The times are chosen so that the varying wind (VW) and constant wind (CW) simulations have the same amount of gas mass in the disk.  In the RC simulations, it is clear that the radius is larger in the CW runs than in the VW run, while in the NC simulations the radius is very similar in the VW and CW runs. }\label{fig-gasradii}
\end{figure}

We first focus on the runs that include radiative cooling (RC runs in cyan and magenta), and it is clear that stripping proceeds significantly differently depending on the wind profile.  First, the total gas mass lost is dramatically different, with much more gas removed in the CW cases.  However, even the rate of gas removal is quite different depending on the wind profile.  With a constant wind (CW), stripping proceeds quickly, and most of the gas removal has occurred less than 1 Gyr after the initial onslaught.  On the other hand, stripping proceeds slowly with a VW.  Even in the last Gyr of the VW runs, when we hold the ram pressure constant at a value that is the same or higher than the CW(L/D) runs, the mass loss rate is lower than in the first few 100 Myrs post-wind of the CW runs.  The amount of gas lost once the VW wind has reached maximum (at $\sim$3 Gyr in the simulation) corresponds to the amount of gas lost in the first $\sim$200 Myr post-wind in the CW runs, and is different in the face-on and angled runs.  In fact, the total gas mass lost after 3 Gyr of stripping (total simulation time of 4 Gyr) is less than the gas mass removed after 500 Myr in the CW cases.  We highlight that even with a lower ram pressure, the RCFOCWL galaxy quickly loses much more gas than the RCFOVW galaxy over the length of the simulation.  Clearly, an increasing ram pressure strips less gas than a constant ram pressure in a radiatively cooled disk.  

We have also run a simulation that has been allowed to radiatively cool and form dense clumps for an extra 2 Gyr before being hit with a constant wind (RCFOCWD).  Indeed, the original thermal pressure surrounding this galaxy is higher than the ram pressure experienced by RCFOVW until $\sim$2.25 Gyr into the varying-wind simulation (1.25 Gyr after the wind hits the galaxy), and the thermal pressure of the stripping wind in RCFOCWD  is also higher than that in RCFOVW.  Despite this, the gas disk mass evolves in a similar fashion to RCFOCW(L).  This indicates that it is not merely time gas is allowed to cool in the simulation, or the surrounding thermal pressure, but the profile of the ram pressure impacting the galaxy that causes the different gas mass loss rates.

We highlight that although much of this paper is focused on simulations with a face-on wind, the differences hold for galaxies inclined to the wind direction (RCVW compared to RCCW).  Because most ram pressure stripped galaxies have some inclination with respect to the ICM wind, verifying these results is important.  The CW runs reflect the results in Roediger \& Bruggen (2005), that higher inclination angles have slightly less gas stripped.  The disk mass within the VW runs is always within 5\%, indicating that the inclination angle may have even less impact on the gas stripping with increasing ram pressure.

The story is different in the cases without radiative cooling (``NC" runs in green).  In the CW run, gas mass removal continues at a high rate throughout the simulation, although the slope decreases slightly with time.  In the VW run, the rate of gas mass loss increases with time as the ram pressure strength increases.  Indeed, for the last Gyr of the VW run, when the ram pressure is held at the maximum (CW) value, the gas removal rate between the CW and VW runs is very similar.  While the total gas mass loss is higher in the CW run than in the VW run, if we compare the VW and CW runs 1 Gyr after the ram pressure reaches its maximum value (4 Gyr into the VW simulation and 2 Gyr into the CW simulation), the total gas loss in NCFOVW is larger by only $\sim$10\% than the gas loss in NCFOCW.  Without radiative cooling, the peak ram pressure drives the rate of gas stripping with little to no influence from the ram pressure profile.

\begin{figure*}
\includegraphics[scale=.98,trim=33mm 123mm 0mm 40mm,clip]{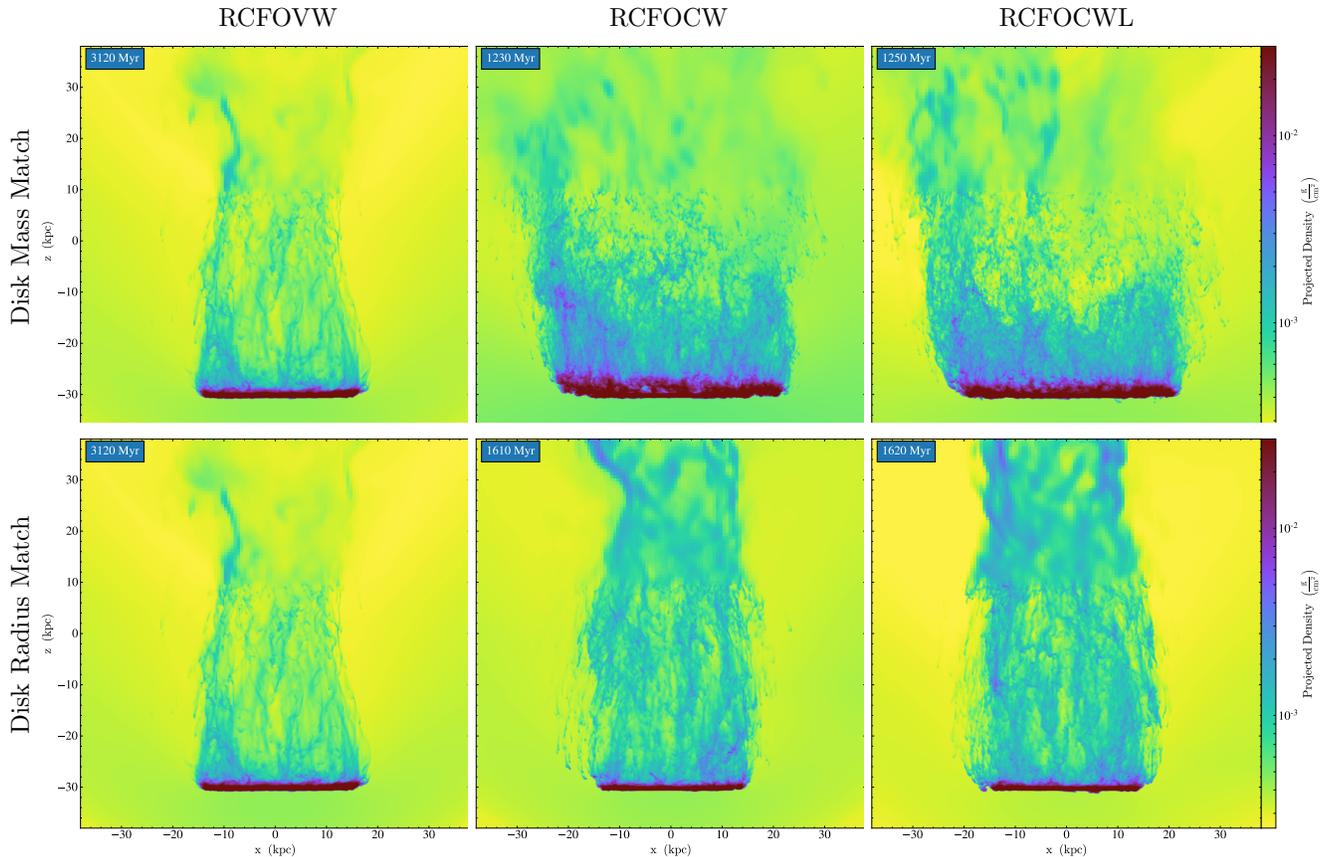}
\caption{Projections of gas density in the RC simulations.  The top panels compare the density projections when the gas masses in the disks are the same, and the bottom panels compare when the gas disk radii are the same. When the mass in the disks agree, the gas in the CW tails is concentrated closer to the disk.  At later times in the CW runs, when the gas disk radius agrees with the VW run, the gas in the tail is more evenly distributed throughout the tail.  See discussion in Section \ref{sec:tailRC} and Figure \ref{fig-rctailmass}}\label{fig-rctailproj}
\end{figure*}

Varying the ram pressure strength has a significant impact on the total gas loss of ram pressure stripped galaxies that include radiative cooling.

\begin{figure}
\includegraphics[scale=0.4]{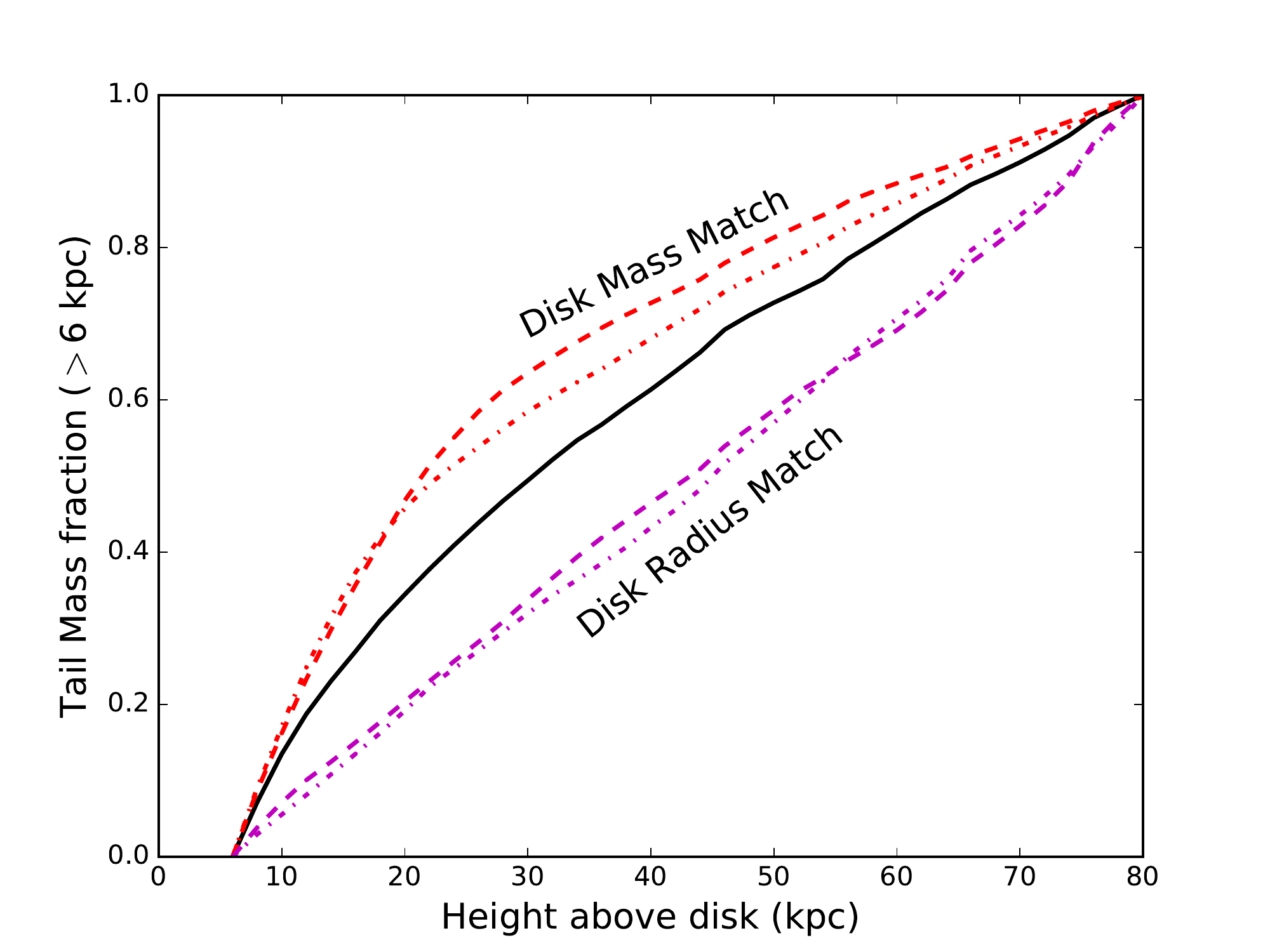}\\
\caption{The cumulative distribution of tail gas along the wind direction in the RC simulations.  The sum begins 6 kpc above the disk and ends at the edge of the box.  The black line is the distribution along the RCFOVW tail at 3.12 Gyr.  The dashed (dash-dotted) lines show the gas distribution in the RCFOCW(L) tails when the disk gas mass or radius is the same (red or magenta).  The gas distribution of the VW and CW tails does not agree at either of these sets of times. }\label{fig-rctailmass}
\end{figure}

\begin{figure*}
\includegraphics[scale=0.945,trim=38mm 123mm 1mm 40mm,clip]{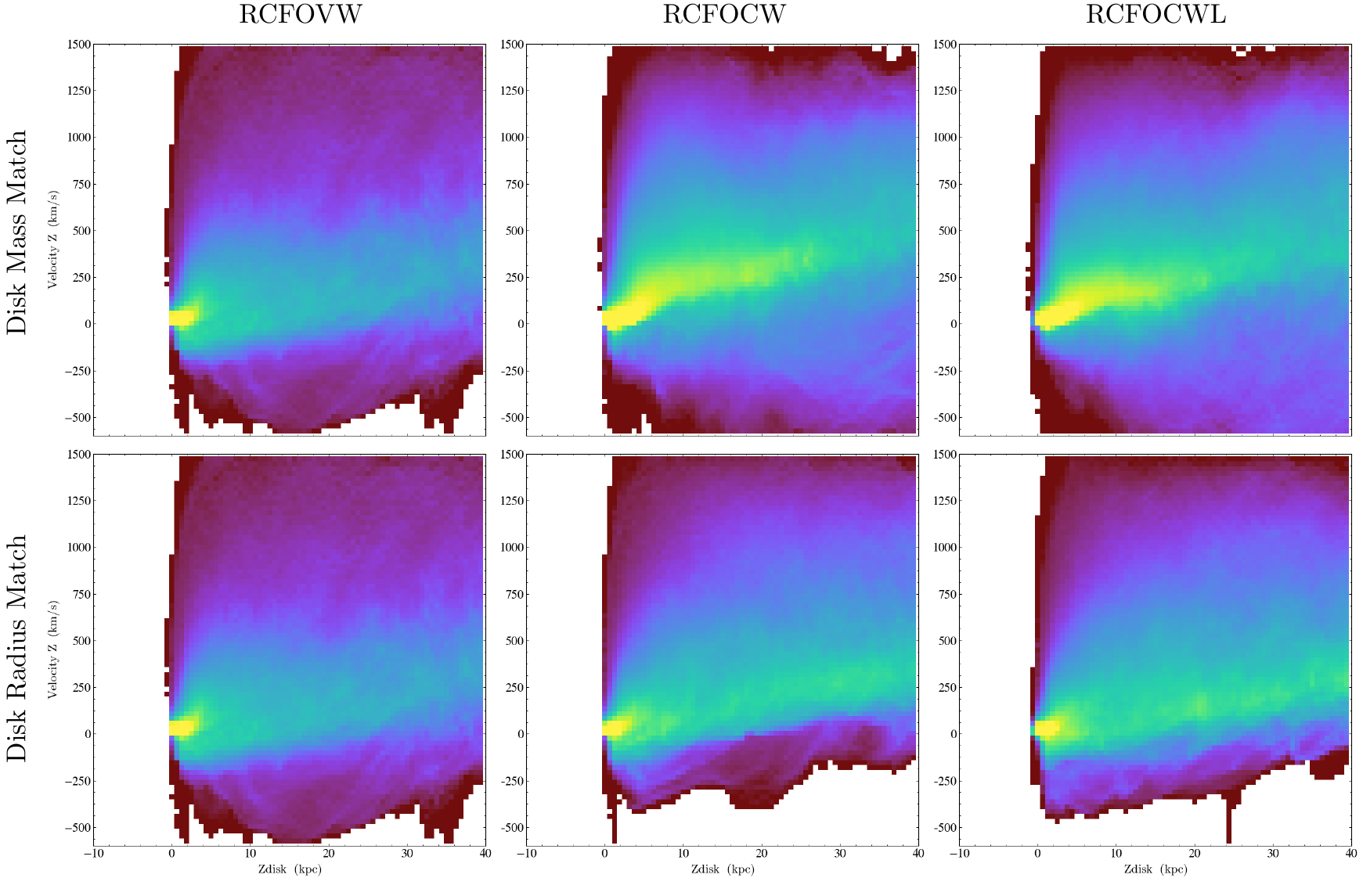}
\includegraphics[scale=0.4,trim=93mm 20.5mm 94mm 4mm,clip]{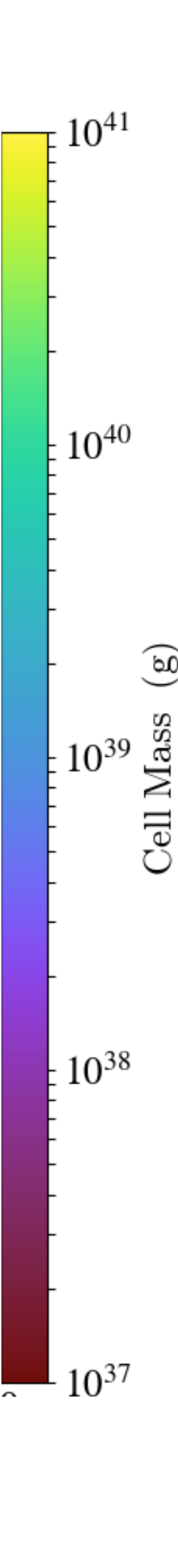} 
\caption{Gas mass distribution as a function of height above the disk and velocity in the wind (z) direction.  The colorbar range is chosen to highlight tail gas at the expense of losing some contrast in gas moving slowly near the disk.  The differing ram pressure profiles is evidenced in the differing velocities along the tail--in the CW runs the tails have higher velocities at larger distances from the galaxy.}\label{fig-rctailvel}
\end{figure*}

\subsection{Gas Profile}\label{sec:gasradius}

We next consider the gas profile in the galaxy disk, focusing on the FO runs for clarity.  This provides a clear comparison between simulations and observations:  the extent of the gas disk.  In addition, we can compare the radial gas density distribution in the disks.  We choose the output from the VW simulations at which the wind at the galaxy midplane has reached its peak value, 3.12 Gyr into the simulation.  We then compare the VW case at 3.12 Gyr to the CW runs when they have the same amount of gas mass in the disk (RCFOCW at 1.23 Gyr, RCFOCWL at 1.25 Gyr, RCFOCWD at 3.24 Gyr, and NCFOCW at 1.58 Gyr).

In Figure \ref{fig-gasradii} we show the azimuthally-average disk gas profile measured using all of the disk gas within $\pm$2 kpc of the disk plane.  We compare the gas profiles of galaxies when the gas mass is the same in the VW and CW runs (within 0.5\%).  Clearly, in the RC simulations, the disk mass does not determine the gas radius.  However, in the NC simulations, the gas radius is quite similar at the times at which the VW and CW runs have the same gas mass (within 0.6\%).  

Examining the gas density profile as a function of galaxy radius, we see that in the RC simulations the galaxies in the CW runs have lower average density in the inner regions than the VW run.  This is despite the fact that we have made sure to select times at which the surrounding ram pressures are the same (or \textit{lower} in the RCFOCWL run).  This is because the initial lower ram pressure that the RCFOVW run experiences cannot remove all but the outermost gas, so instead compresses it so that it more quickly radiatively cools into dense structures.  At later times, when the ram pressure reaches its peak value, the more central gas is too dense to be quickly removed.  This is in contrast to the CW runs, in which the strong ram pressure is able to remove much of the low density gas before compressing it.  The compression of the disk gas in RCFOVW is a continuous process as the ram pressure increases, as evidenced by the fact that stripping in RCFOCWD proceeds quickly starting at 3.2 Gyr into the simulation despite the fact that the gas disk has been allowed to cool into clumps in a relatively high-pressure static ICM.  RCFOCWD has a radial gas density profile much more similar to RCFOCW(L) than to RCFOVW.  While a compression wave still runs through the disk in the NC runs, because they cannot radiatively cool, there is no long-term density increase (see Figure 16 in Tonnesen \& Stone 2014).  We discuss this in detail below in Section \ref{sec:discussion}.

\subsection{Gas Tail}\label{sec:tail_results}

Current comparisons between observations and simulations demand agreement between both disk and tail properties to use simulations to interpret observations.  The gas tail can give important clues to a galaxy's stripping or interaction history, both in the distribution of gas along the direction of motion (in the wind tunnel simulation this is the wind direction) or in the plane of the sky, and in the velocity distribution of gas in the tail.  For example, in Merluzzi et al. (2016), simulations with a constant wind are not able to simultaneously reproduce both the extent of the tail and the truncation of the gas disk for one of the observed galaxies, and thus the authors conclude that ram pressure stripping alone cannot account for the observed gas removal.  Therefore, in order to use a CW simulation in place of one that includes a varying wind from a galaxy's orbit, both disk and tail observables must agree.  

In this section, we compare the tail properties of the VW case at 3.12 Gyr to the CW runs at two times:  when they have the same gas mass in the disk (RCFOCW at 1.23 Gyr, RCFOCWL at 1.25 Gyr, and NCFOCW at 1.58 Gyr), and when they have the same gas disk radius (RCFOCW at 1.61 Gyr, RCFOCWL at 1.62 Gyr, and NCFOCW at 1.58 Gyr).  Because RCFOCWD is so similar to the RCFOCW(L) runs, with qualitatively identical comparisons with RCFOVW, we do not discuss it here.

\subsubsection{Radiative Cooling Runs}\label{sec:tailRC}

Before we focus on the tail properties, it is worthwhile to recall that the CW runs have been stripped for relatively short amounts of time when the gas disk masses and radii agree with the VW run at 3.12 Gyr.  In fact, when the CW galaxies have the same gas disk mass as the VW galaxy, stripping has just begun ($\sim$100 Myr) and the galaxies will continue to quickly lose gas mass for another $\sim$400 Myr.  When the gas disk radii agree, gas removal from the CW galaxies has begun to slow, and we are looking at a later stripping stage. 

In Figure \ref{fig-rctailproj} we compare the density projections of RCFOVW with the RCFOCW and RCFOCWL runs.  The top panels compare when the gas masses within $\pm$ 10 kpc of the disk plane are the same, and the bottom panels compare when the disk radii measured using the gas profiles are the same.  We clearly see that the gas density distribution in the VW case differs from the CW cases.  When the gas mass agrees across the runs, the tails in the CW runs are broader, reflecting the larger surviving disk (see Figure \ref{fig-gasradii}).  In the CW tails there is more high density gas closer to the disk (within 20 kpc), which may reflect the fact that the wind has been stripping the CW galaxies for only ~100 Myr, in comparison to the ~2 Gyr of stripping in the VW run.  There has not been enough time in the CW runs for stripped gas, particularly high-density gas that moves more slowly away from the disk (Tonnesen \& Bryan 2010), to reach large distances.  However, we do note that all three tails have gas that is denser than the surrounding ICM extending to the edge of the projection.

When the gas radii agree across the runs (the bottom panels), we see that there is more high-density gas throughout the CW tails.  This is ~400 Myr later in the CW runs, during which time gas has continued to be removed from the disk at a relatively constant high rate (Figure \ref{fig-diskmass}), so we would expect more gas distributed throughout the tail.  However, from the projections it is difficult to determine whether the gas distribution is different or if the higher density simply reflects that there is more gas in the CW tails when the disk radii are the same.  

We examine the gas tail quantitatively in Figure \ref{fig-rctailmass}, which plots the cumulative mass distribution of gas in the tail along the wind direction starting at 6 kpc above the disk plane out to the edge of the simulated box (the lower boundary choice has no qualitative effect on the results).  When the disk mass is the same in the VW and CW runs, the bulk of the stripped gas is found closer to the CW(L) disks.  Conversely, when the VW and CW runs have the same gas disk radii, the tail mass in the CW runs is more evenly-distributed along the tail, with more mass farther from the disk. 

We next consider whether the different tail gas distributions are reflected in the tail gas velocity, which is another observed property of stripped galaxies.  In Figure \ref{fig-rctailvel} we plot the distribution of gas mass with a tracer fraction of at least 0.25 as a function of distance above the disk and velocity in the wind direction.  Here we only focus on the refined region of the box, within 40 kpc of the disk, as this region better resolves the velocity structure in the tail.  When the gas mass agrees, we first see that there is more gas close to the disk.  In addition, much of the stripped gas in the CW tails is accelerated to higher velocities within 4-10 kpc than in the VW tail.  While this is most dramatic in the RCFOCW run, it can also be seen in the RCFOCWL run even though the ram pressure strength is lower than in the RCFOVW panel.  The faster velocity of tail gas continues to the edge of the refined region, 40 kpc from the disk.  Also, the CW runs have a broader velocity distribution that extends to much higher ($\sim$2x) velocities when focusing on the higher-mass contours ($\ge$10$^{39}$ g).  

In the bottom panels, we see that when the gas disk radius agrees well between the VW and CW cases, the flow of most of the gas in the tail is also similar.  However, there tends to be more fallback in the VW run, and more gas moving at high velocities in the CW runs.  This may be because the ram pressure has been stronger and the wind faster for longer in the CW runs, even when the maximum wind velocity is lower (in RCFOCWL).  We will discuss this in more detail in Section \ref{sec:discussion}.

Neither the gas density nor the velocity distribution in the tails agree when comparing radiatively cooled stripped galaxies in simulations with constant and varying ICM winds.  

\subsubsection{No Cooling Runs}

As we discussed in Section \ref{sec:gasradius}, without radiative cooling the disk masses and radii of the VW and CW simulations both agree using a single pair of outputs (VW at 3.12 Gyr, CW at 1.58 Gyr).  We now compare the stripped tails at those outputs.

\begin{figure}
\begin{center}
\includegraphics[scale=0.27]{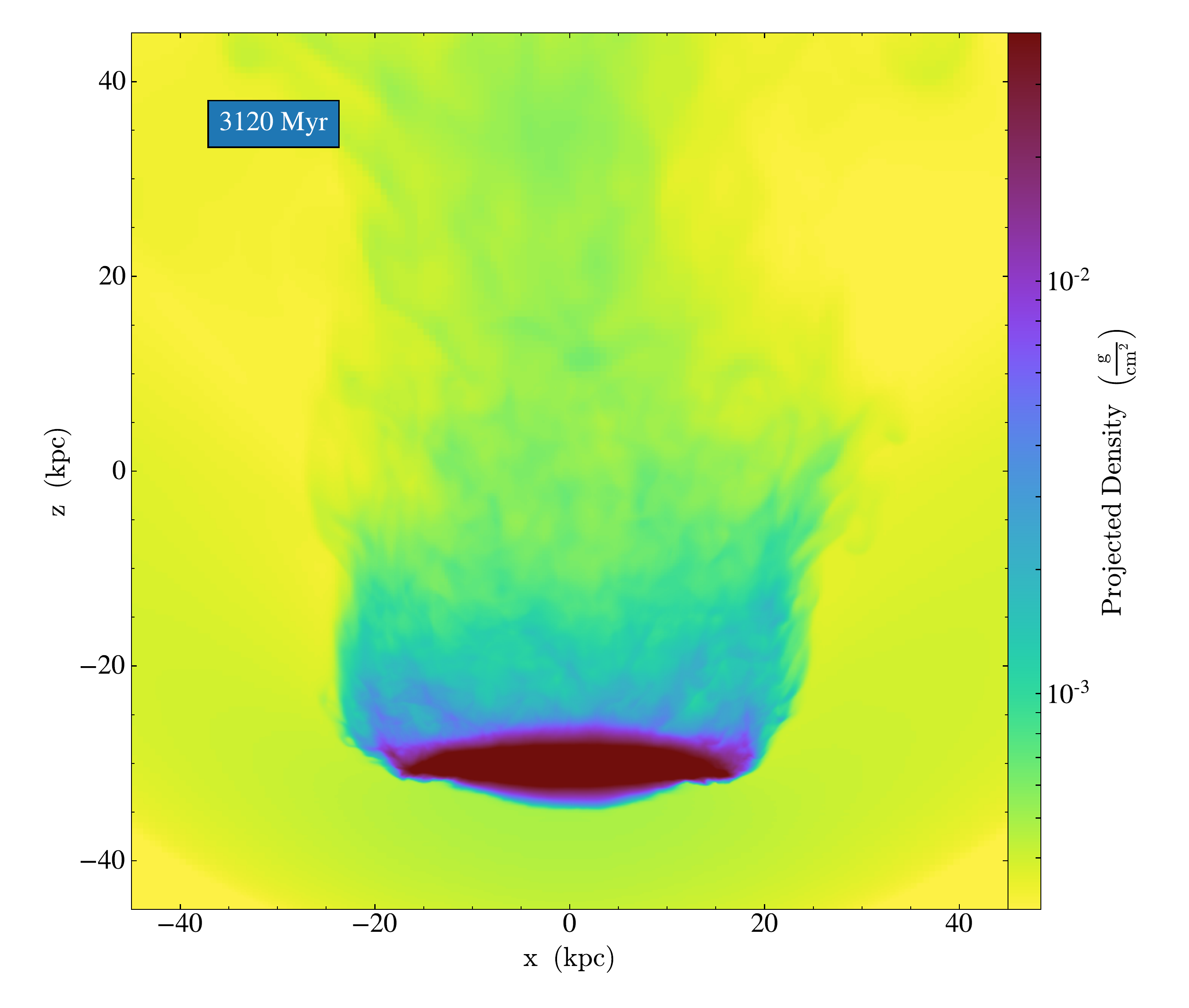}\\
\includegraphics[scale=0.27]{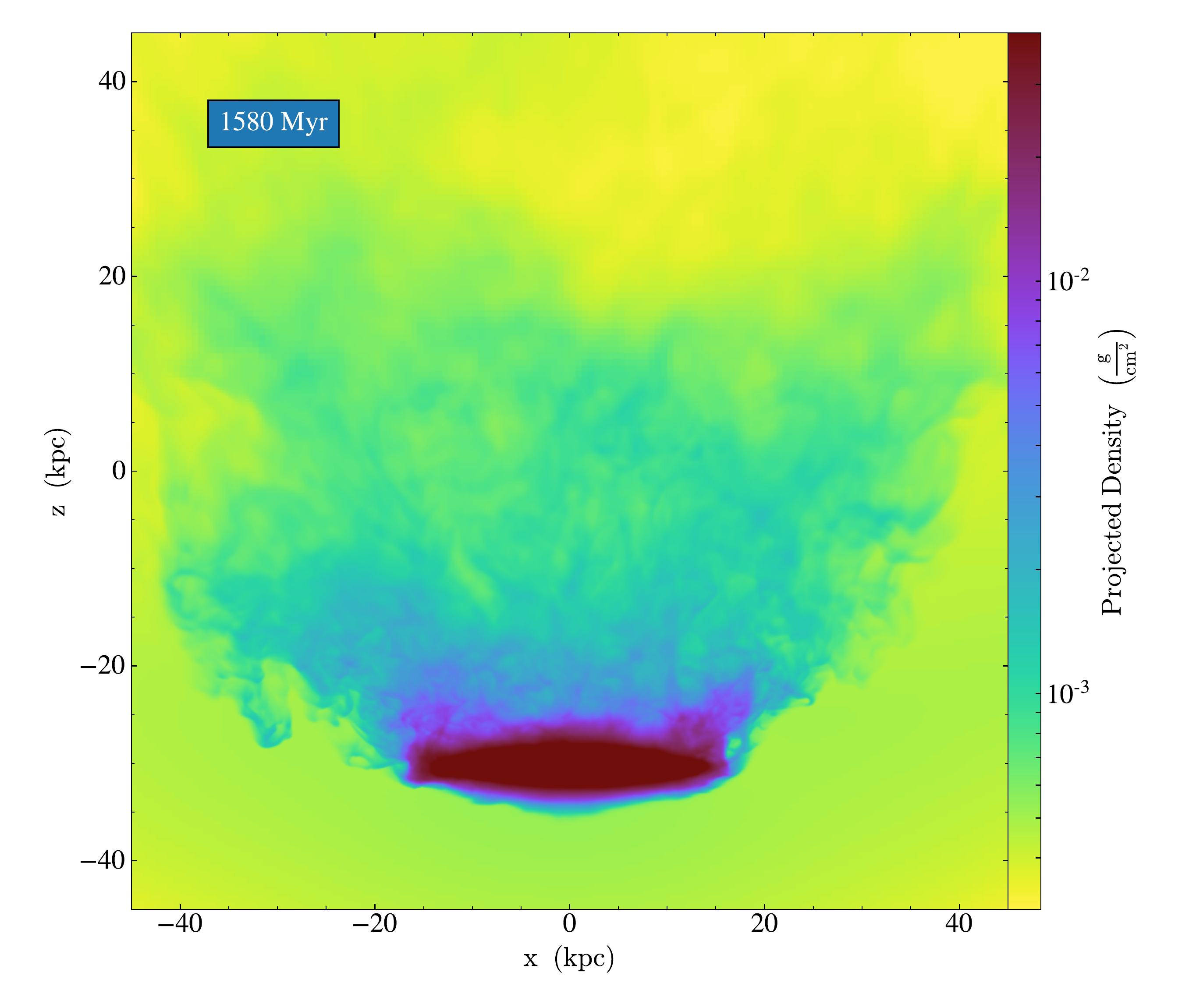}\\
\caption{Projections of gas density in the NC simulations.  The top panel shows NCFOVW 3.12 Gyr into the simulation, and the bottom panel shows NCFOCW when the disk properties (gas mass and radius) agree with the NCFOVW run.  The gas distribution in the projection is quite different in the VW and CW tails.}\label{fig-nctailproj}
\end{center}
\end{figure}

\begin{figure}
\begin{center}
\includegraphics[scale=0.95,trim=18mm 132mm 133mm 22mm,clip]{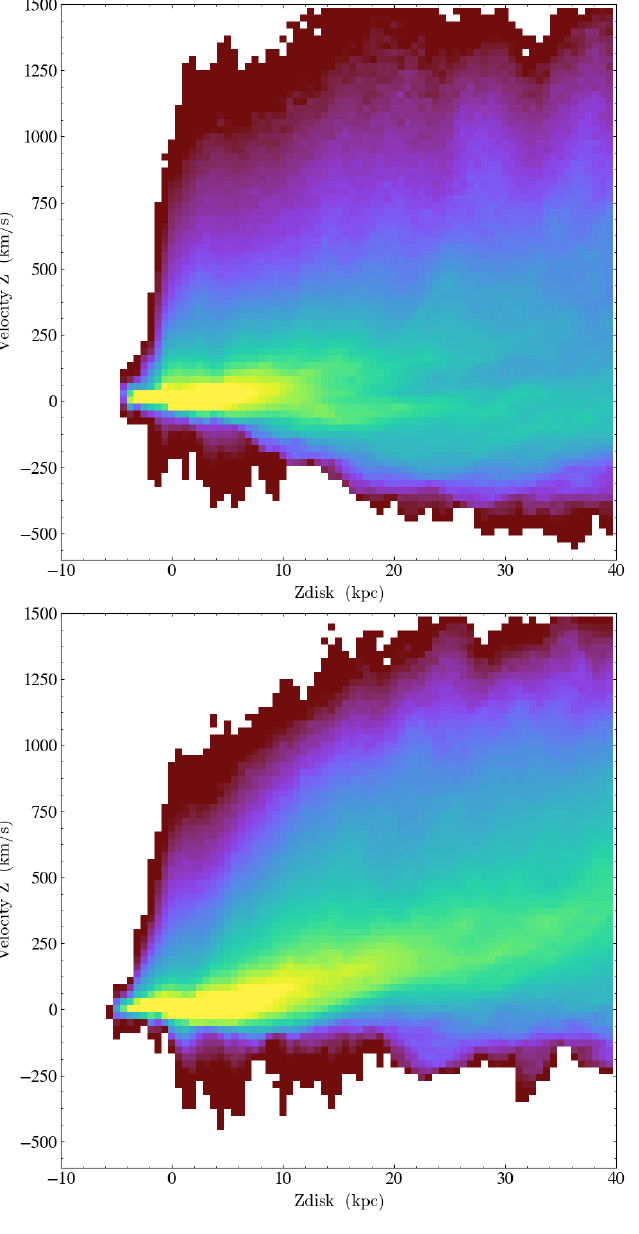}
\includegraphics[scale=0.45,trim=93mm 2.5mm 94mm 4mm,clip]{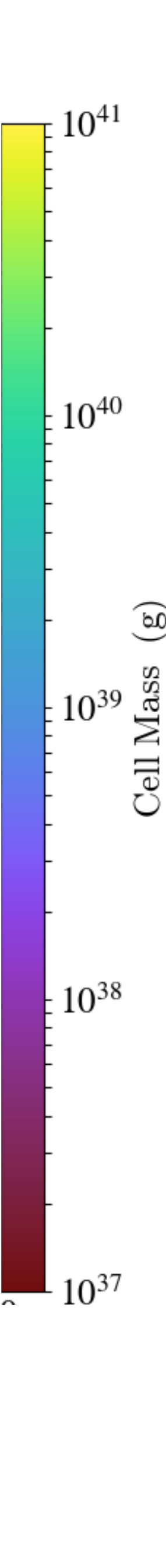}
\caption{Gas mass distribution as a function of height above the disk and velocity in the wind (z) direction.  The top and bottom panels show the NCFOVW and NCFOCW runs, respectively.  As in the RC simulations, the stripped tail has a higher velocity far from the galaxy in the CW run compared to the VW run.}\label{fig-nctailvel}
\end{center}
\end{figure}

In Figure \ref{fig-nctailproj} we show density projections of the simulations, as in Figure \ref{fig-rctailproj}.  Clearly the morphology of these tails are quite different, with the CW run having a much more flared tail.  In the CW tail, the high density gas near the disk is more evenly distributed in projected radius, while in the VW tail the higher density gas tends to be found at larger cylindrical radii.  Also, more higher density gas is found in the CW tail at all heights above the disk starting at about 4 kpc.  The cumulative mass distributions of gas in the tail along the wind direction are quite similar, but the differing density distributions of gas in the VW and CW tails means that these tails look quite different.

Finally, we consider the velocity distribution of gas in the tail.  In Figure \ref{fig-nctailvel} we plot the distribution of gas mass with a tracer fraction of at least 0.25 as a function of distance above the disk and velocity in the wind direction.  As in Figure \ref{fig-rctailvel}, we only focus on the refined region of the box, within 40 kpc of the disk.  The tail in the VW run has a significant component with negative velocities that stretches at least 40 kpc from the disk, while very little gas is falling back in the CW run.  In the CW run much more gas is moving at high velocities, possibly because it has been affected by a fast-moving ICM for longer than the VW run (as in the RC simulations).  

As with the simulations with radiative cooling, neither the gas density nor the velocity distribution in the tails agree when comparing stripped galaxies in non-cooling simulations with constant and varying ICM winds.  

\section{Comparing Simulations with and without Radiative Cooling}\label{sec:discussion}

In this paper, we have run simulations that include radiative cooling (RC) and those that do not (NC).  While clearly radiative cooling is an important process occurring in ram pressure stripped galaxies like those we simulate, we know that we are not including several important physical processes.  For example, we are only including radiative cooling down to $\sim$10$^4$ K.  We do not include star formation and subsequent energy input from supernovae and stellar feedback.  We also ignore heating from the UV background, cosmic rays, or the surrounding ICM.  Turbulent scales below our resolution are not considered, and we do not include magnetic fields.  

It is useful to highlight how the subgrid physics we implement in our simulations may affect our results as we draw conclusions and compare to observations.  Therefore, here we consider the different results in our RC and NC simulations.  One main difference is that the RCFOVW and RCFOCW(L/D) runs have different stripping rates and gas profiles, while in the NCFOVW and NCFOCW the disks are quite similar (Figures \ref{fig-diskmass} \& \ref{fig-gasradii}).  Radiative cooling allows the disk to collapse and dense clouds to form even before the wind hits the disk, (seen at later times by comparing Figures \ref{fig-rctailproj} and \ref{fig-nctailproj}).

When the ICM wind impacts the galaxy, it compresses gas in the galaxy, as noted in Sections \ref{sec:disk_results} \& \ref{sec:gasradius}.  This compression wave can be important to the evolution of the galaxy with radiative cooling.  In the VW run, because the initial ram pressure is quite low, this results in radiative cooling of gas that was previously too low density to quickly cool but cannot yet be removed.  By the time the peak ram pressure is impacting the galaxy in the varying wind run, more gas is too dense to be stripped.  In Figure \ref{fig-gasradii} we see that the inner density profile of RCFOVW has higher density gas than the RCFOCW(L/D) runs.  This is also the case when comparing the gas profiles at the times when the gas radii agree.  However, before the wind hits the disks, the inner density of the RCFOVW galaxy is less than in the RCFOCW(L/D) galaxies, supporting the picture of dominant cooling in RCFOVW versus dominant stripping in RCFOCW(L/D).  Because the initial surrounding thermal pressure in RCFOCWD is higher than the ram pressure experienced by RCFOVW until $\sim$2.25 Gyr into the varying-wind simulation (1.25 Gyr after the wind hits the galaxy), see Figure \ref{fig-winds}, the additional compression of the disk gas in the RCFOVW simulation due to the increasing ram pressure is important to the gas stripping.

While a compression wave is likely to drive cooling and collapse of gas in a disk, we are not including heating sources that could mitigate the effect of cooling.  Also, without star formation or feedback, dense clouds survive for a significant period of time (the entire simulation unless destroyed by the wind), which may allow more time for the clouds to travel towards the center of the disk and thus factor into the different density profiles (e.g. Schulz \& Struck 2001; Tonnesen \& Bryan 2009; 2012).  Despite these caveats, the differences in the disk properties between the RCFOVW and RCFOCW(L) runs is enough to question the detailed comparison of constant-wind simulations with observations for any individual galaxy.  

Although the level of agreement between disks in the VW and CW runs depends on radiative cooling, the results regarding the stripped tail are independent of whether cooling is implemented.  A simulation using a constant wind will not reproduce the tail properties of a simulation with a varying wind.  When  disk properties agree, although the total time that the constant wind has been stripping the galaxy is shorter than that of the varying wind, the constant wind has had a longer time at the peak velocity.  This allows some of the gas to be accelerated to higher velocities, a phenomenon we point out for both the RC and NC runs in Figures \ref{fig-rctailvel} and \ref{fig-nctailvel}.  We also note that any length of time at a constant ram pressure is unphysical for nearly all galaxy orbits.

Finding agreement in disk properties between simulations with constant and varying winds does not indicate that tail properties, seen most strongly in the velocity profiles, will agree.  This robust result does not depend on the exact physics implemented in a simulation.  This is well understood in terms of the multi-stage stripping described by, e.g. Schulz \& Struck (2001) and Roediger \& Hensler (2005).  Initially gas is removed from the disk, but remains bound to the galaxy and may ``hang" behind the galaxy, especially when it is shielded from the ICM wind by the remaining gas disk.  With continued ram pressure, most gas is this region is eventually stripped, but some gas falls back towards the disk.  Roediger \& Hensler (2005) show that stronger ram pressure decreases both the amount of time that gas remains bound in the halo and the fraction of gas that falls back to the disk.  For our simulations, this means that because the VW runs spend most of the simulation before our comparison time at 3.12 Gyr with ram pressure well below those of the CW run, more gas will hang for longer in the halo, leading to slower velocities throughout the VW tail.

Another dramatic difference in the velocity of tail gas is the fact that there is more fallback in the VW runs, as seen in both the RC and NC runs in Figures \ref{fig-rctailvel} \& \ref{fig-nctailvel}.  This may be because the snapshots of the VW runs are taken $\sim$2 Gyr after the wind has hit the disk, while the comparison times of the CW runs are no more than 600 Myr after the wind has hit the disk.  As discussed in Roediger et al. (2015), after an impulsive onset of a wind there can be a several-hundred Myr relaxation phase (depending on the size of the object and the flow velocity) before stripping reaches a quasi-steady state in which a backflow develops behind the unstripped gas.  The VW runs should be well into the quasi-steady stripping phase, while the CW runs may not yet have developed the backflow velocity structure.  Indeed, if we consider the velocity of stripped gas in the CW runs at 3.12 Gyr we see very similar levels of fallback. 

The rapidly shrinking disk in the radiatively cooled CW runs exacerbates the difference in the tail velocities.  As pointed out in Roediger et al. (2015), as the shielding region shrinks, more gas can be directly accelerated by the ICM wind.  In the radiatively cooled simulations, the disk radius of the CW runs continues shrinking for about 1 Gyr after the wind hits the disk, while the radius of the gas disk in the VW run changes very little after about 2.5 Gyr into the simulation, before the ram pressure reaches its peak.  Therefore, when we compare the tails, the CW disks are shrinking and more gas is being pushed from behind the galaxies while the VW disk size remains relatively constant.

\subsection{Resolution}

While we do not perform a resolution test in this series of simulations, previous work has discussed the effects of resolution at length.  For example, Roediger \& Bruggen (2006) found that changing resolution has very little effect on wind-tunnel simulations without radiative cooling.  Tonnesen \& Bryan (2009) found that in wind-tunnel simulations with radiative cooling, lower spatial resolution runs have larger cold clouds with lower maximum densities.  This means a smoother disk and stripping rates closer to those of disks with no radiative cooling.  Tonnesen \& Bryan (2010) considered the effects of lower resolution on stripped tails with radiative cooling, and found that less resolution results in shorter tails, both from easier mixing of stripped gas into the ICM and from less low-density gas accelerated quickly from the disk.  They also found more fallback in lower resolution runs, possibly because the remaining disk shields the stripped gas more effectively with less fragmentation.   

Therefore, we predict that lowering the resolution of the radiatively cooled runs would weaken our results, but only with very low resolution would we reach the results of the no-cooling runs.

\section{Conclusions}\label{sec:conclusions}

In this paper we examine a series of hydrodynamical simulations to determine whether considering the varying ram pressure strength due to the orbit of a satellite galaxy is important when using simulations to interpret observations.  Specifically, we compare simulations with increasing ram pressure to those with a constant ram pressure.  Our main results are as follows:

1)  In simulations that include radiative cooling, the amount of gas removed from the disk and rate of removal is dramatically different in a constant versus a varying wind.  More gas is removed more quickly from a constant wind.  Even once the varying wind reaches the peak ram pressure, the gas removal rate is lower than in the constant wind simulations (Figure \ref{fig-diskmass}).  

2)  In radiative cooling simulations, when the gas mass is the same in simulations with a constant and varying wind, the radius of the surviving gas disks differ (Fig. \ref{fig-gasradii}). 

3)  The agreement of disk properties, such as gas mass or radius, between simulations with different wind properties does not indicate that the tail properties, such as density or velocity structure, will be similar.  This occurs whether or not radiative cooling is included in the simulations (Section \ref{sec:tail_results}).

It is apparent that a simulation that uses a constant wind cannot accurately reproduce a galaxy that has been ram pressure stripped by an increasing wind.  Importantly, in this paper, we examine simulations that include radiative cooling and those that do not, and find that the cooling implementation does not affect our result.  Thus, we stress that simulations, particularly those with a constant wind, should be used carefully and sparingly as tools to interpret specific observations of galaxies.

Using simulations to interpret observations of individual systems requires sampling a large set of parameters.  The wind angle, ram pressure strength, and gas disk scale height will affect the resulting stripped galaxy's gas density and velocity structure (e.g. Merluzzi et al. 2013, 2016).  In this paper we have shown that including an increasing ram pressure also has an effect.  More complications are certainly possible, for example, a varying wind angle may effect the tail morphology and velocity structure.  Also, a clumpy or turbulent ICM may effect the efficiency of stripping and mixing.  The surrounding gravitational potential, both from the cluster and nearby galaxies, are also likely to affect the stripped tail.  

Testing the effects of more complicated ICM and orbital models on cluster galaxies will not only allow us to test how galaxies are affected by their surroundings, but also to determine the extent to which we can use ram pressure stripped galaxies to probe the nature of the surrounding ICM--its turbulence, pressure, and magnetic field.

\acknowledgements I would like to thank the referee, Elke Roediger, for comments that greatly improved the paper.  I would also like to thank the GASP team for useful discussions about ram pressure stripping, and Greg Bryan for helpful conversations and comments on the paper.  The simulations were run with computing resources provided by mies.  ST was partially supported by the Alvin E. Nashman Fellowship in Theoretical Astrophysics.


\begin{thebibliography}

\bibitem[Bekki(2014)]{2014MNRAS.438..444B} Bekki, K.\ 2014, \mnras, 438, 444 


\bibitem[Bellhouse et al.(2017)]{2017ApJ...844...49B} Bellhouse, C., Jaff{\'e}, Y.~L., Hau, G.~K.~T., et al.\ 2017, \apj, 844, 49 

\bibitem[Boselli et al.(2018)]{2018A&A...615A.114B} Boselli, A., Fossati, M., Cuillandre, J.~C., et al.\ 2018, \aap, 615, A114 

\bibitem[Bovy(2015)]{2015ApJS..216...29B} Bovy, J.\ 2015, \apjs, 216, 29 


\bibitem[Bryan et al.(2014)]{2014ApJS..211...19B} Bryan, G.~L., Norman, M.~L., O'Shea, B.~W., et al.\ 2014, \apjs, 211, 19 


\bibitem[Burkert(1995)]{1995ApJ...447L..25B} Burkert, A.\ 1995, \apjl, 447, L25 


\bibitem[Cayatte et al.(1990)]{1990AJ....100..604C} Cayatte, V., van Gorkom, J.~H., Balkowski, C., \& Kotanyi, C.\ 1990, \aj, 100, 604 


\bibitem[Chung et al.(2009)]{2009AJ....138.1741C} Chung, A., van Gorkom, J.~H., Kenney, J.~D.~P., Crowl, H., \& Vollmer, B.\ 2009, \aj, 138, 1741 


\bibitem[Chung et al.(2007)]{2007ApJ...659L.115C} Chung, A., van Gorkom, J.~H., Kenney, J.~D.~P., \& Vollmer, B.\ 2007, \apjl, 659, L115 


\bibitem[Dressler \& Gunn(1983)]{1983ApJ...270....7D} Dressler, A., \& Gunn, J.~E.\ 1983, \apj, 270, 7 


\bibitem[Evrard(1991)]{1991MNRAS.248P...8E} Evrard, A.~E.\ 1991, \mnras, 248, 8P 


\bibitem[Fossati et al.(2016)]{2016MNRAS.455.2028F} Fossati, M., Fumagalli, M., Boselli, A., et al.\ 2016, \mnras, 455, 2028 


\bibitem[Fujita(1998)]{1998ApJ...509..587F} Fujita, Y.\ 1998, \apj, 509, 587 


\bibitem[Fujita \& Nagashima(1999)]{1999ApJ...516..619F} Fujita, Y., \& Nagashima, M.\ 1999, \apj, 516, 619 


\bibitem[Gavazzi et al.(2006)]{2006A&A...446..839G} Gavazzi, G., Boselli, A., Cortese, L., et al.\ 2006, \aap, 446, 839 

\bibitem[George et al.(2018)]{2018MNRAS.479.4126G} George, K., Poggianti, B.~M., Gullieuszik, M., et al.\ 2018, \mnras, 479, 4126 

\bibitem[Gnedin(2003)]{2003ApJ...589..752G} Gnedin, O.~Y.\ 2003, \apj, 589, 752 


\bibitem[Gullieuszik et al.(2017)]{2017ApJ...846...27G} Gullieuszik, M., Poggianti, B.~M., Moretti, A., et al.\ 2017, \apj, 846, 27 


\bibitem[Gunn \& Gott(1972)]{1972ApJ...176....1G} Gunn, J.~E., \& Gott, J.~R., III 1972, \apj, 176, 1 


\bibitem[Haynes et al.(1984)]{1984ARA&A..22..445H} Haynes, M.~P., Giovanelli, R., \& Chincarini, G.~L.\ 1984, \araa, 22, 445 


\bibitem[Haynes et al.(2007)]{2007ApJ...665L..19H} Haynes, M.~P., Giovanelli, R., \& Kent, B.~R.\ 2007, \apjl, 665, L19 


\bibitem[Hernquist(1993)]{1993ApJS...86..389H} Hernquist, L.\ 1993, \apjs, 86, 389 


\bibitem[Irwin \& Sarazin(1996)]{1996ApJ...471..683I} Irwin, J.~A., \& Sarazin, C.~L.\ 1996, \apj, 471, 683 

\bibitem[J{\'a}chym et al.(2009)]{2009A&A...500..693J} J{\'a}chym, P., K{\"o}ppen, J., Palou{\v s}, J., \& Combes, F.\ 2009, \aap, 500, 693 


\bibitem[J{\'a}chym et al.(2007)]{2007A&A...472....5J} J{\'a}chym, P., Palou{\v s}, J., K{\"o}ppen, J., \& Combes, F.\ 2007, \aap, 472, 5 


\bibitem[J{\'a}chym et al.(2014)]{2014ApJ...792...11J} J{\'a}chym, P., Combes, F., Cortese, L., Sun, M., \& Kenney, J.~D.~P.\ 2014, \apj, 792, 11 


\bibitem[J{\'a}chym et al.(2017)]{2017ApJ...839..114J} J{\'a}chym, P., Sun, M., Kenney, J.~D.~P., et al.\ 2017, \apj, 839, 114 


\bibitem[Kapferer et al.(2009)]{2009A&A...499...87K} Kapferer, W., Sluka, C., Schindler, S., Ferrari, C., \& Ziegler, B.\ 2009, \aap, 499, 87 


\bibitem[Kenney et al.(2008)]{2008ApJ...687L..69K} Kenney, J.~D.~P., Tal, T., Crowl, H.~H., Feldmeier, J., \& Jacoby, G.~H.\ 2008, \apjl, 687, L69 


\bibitem[Koopmann \& Kenney(2004)]{2004ApJ...613..866K} Koopmann, R.~A., \& Kenney, J.~D.~P.\ 2004, \apj, 613, 866 


\bibitem[Larson et al.(1980)]{1980ApJ...237..692L} Larson, R.~B., Tinsley, B.~M., \& Caldwell, C.~N.\ 1980, \apj, 237, 692 

\bibitem[Lee \& Chung(2018)]{2018ApJ...866L..10L} Lee, B., \& Chung, A.\ 2018, \apjl, 866, L10 


\bibitem[Mandelbaum et al.(2008)]{2008JCAP...08..006M} Mandelbaum, R., Seljak, U., \& Hirata, C.~M.\ 2008, Journal of Cosmology and Astroparticle Physics, 8, 006 


\bibitem[Merluzzi et al.(2016)]{2016MNRAS.460.3345M} Merluzzi, P., Busarello, G., Dopita, M.~A., et al.\ 2016, \mnras, 460, 3345 


\bibitem[Merluzzi et al.(2013)]{2013MNRAS.429.1747M} Merluzzi, P., Busarello, G., Dopita, M.~A., et al.\ 2013, \mnras, 429, 1747 


\bibitem[Merritt(1984)]{1984ApJ...276...26M} Merritt, D.\ 1984, \apj, 276, 26 


\bibitem[Miyamoto \& Nagai(1975)]{1975PASJ...27..533M} Miyamoto, M., \& Nagai, R.\ 1975, \pasj, 27, 533 


\bibitem[Moore et al.(1996)]{1996Natur.379..613M} Moore, B., Katz, N., Lake, G., Dressler, A., \& Oemler, A.\ 1996, \nat, 379, 613 


\bibitem[Moran et al.(2007)]{2007ApJ...671.1503M} Moran, S.~M., Ellis, R.~S., Treu, T., et al.\ 2007, \apj, 671, 1503 


\bibitem[Moretti et al.(2018)]{2018MNRAS.475.4055M} Moretti, A., Poggianti, B.~M., Gullieuszik, M., et al.\ 2018a, \mnras, 475, 4055 

\bibitem[Moretti et al.(2018)]{2018MNRAS.480.2508M} Moretti, A., Paladino, R., Poggianti, B.~M., et al.\ 2018b, \mnras, 480, 2508 


\bibitem[Mori \& Burkert(2000)]{2000ApJ...538..559M} Mori, M., \& Burkert, A.\ 2000, \apj, 538, 559 


\bibitem[Oosterloo \& van Gorkom(2005)]{2005A&A...437L..19O} Oosterloo, T., \& van Gorkom, J.\ 2005, \aap, 437, L19 


\bibitem[Poggianti et al.(2016)]{2016AJ....151...78P} Poggianti, B.~M., Fasano, G., Omizzolo, A., et al.\ 2016, \aj, 151, 78 


\bibitem[Poggianti et al.(2009)]{2009ApJ...693..112P} Poggianti, B.~M., Arag{\'o}n-Salamanca, A., Zaritsky, D., et al.\ 2009, \apj, 693, 112 


\bibitem[Poggianti et al.(2004)]{2004ApJ...601..197P} Poggianti, B.~M., Bridges, T.~J., Komiyama, Y., et al.\ 2004, \apj, 601, 197 


\bibitem[Poggianti et al.(2017)]{2017ApJ...844...48P} Poggianti, B.~M., Moretti, A., Gullieuszik, M., et al.\ 2017, \apj, 844, 48 

\bibitem[Poggianti et al.(2019)]{2019MNRAS.482.4466P} Poggianti, B.~M., Gullieuszik, M., Tonnesen, S., et al.\ 2019, \mnras, 482, 4466 

\bibitem[Roediger \& Br{\"u}ggen(2008)]{2008MNRAS.388..465R} Roediger, E., \& Br{\"u}ggen, M.\ 2008, \mnras, 388, 465 


\bibitem[Roediger \& Br{\"u}ggen(2007)]{2007MNRAS.380.1399R} Roediger, E., \& Br{\"u}ggen, M.\ 2007, \mnras, 380, 1399 



\bibitem[Roediger \& Br{\"u}ggen(2006)]{2006MNRAS.369..567R} Roediger, E., \& Br{\"u}ggen, M.\ 2006, \mnras, 369, 567 

\bibitem[Roediger \& Hensler(2005)]{2005A&A...433..875R} Roediger, E., \& Hensler, G.\ 2005, \aap, 433, 875 


\bibitem[Roediger et al.(2015)]{2015ApJ...806..103R} Roediger, E., Kraft, R.~P., Nulsen, P.~E.~J., et al.\ 2015, \apj, 806, 103 


\bibitem[Roediger et al.(2015)]{2015ApJ...806..104R} Roediger, E., Kraft, R.~P., Nulsen, P.~E.~J., et al.\ 2015, \apj, 806, 104 



\bibitem[Sarazin \& White(1987)]{1987ApJ...320...32S} Sarazin, C.~L., \& White, R.~E., III 1987, \apj, 320, 32 


\bibitem[Schulz \& Struck(2001)]{2001MNRAS.328..185S} Schulz, S., \& Struck, C.\ 2001, \mnras, 328, 185 


\bibitem[Smith et al.(2010)]{2010A&A...518L..18S} Smith, G.~P., Haines, C.~P., Pereira, M.~J., et al.\ 2010, \aap, 518, L18 


\bibitem[Smith et al.(2012)]{2012MNRAS.420.1990S} Smith, R., Fellhauer, M., \& Assmann, P.\ 2012, \mnras, 420, 1990 


\bibitem[Sun et al.(2006)]{2006ApJ...637L..81S} Sun, M., Jones, C., Forman, W., et al.\ 2006, \apjl, 637, L81 

\bibitem[Takeda et al.(1984)]{1984MNRAS.208..261T} Takeda, H., Nulsen, P.~E.~J., \& Fabian, A.~C.\ 1984, \mnras, 208, 261 


\bibitem[Tasker \& Bryan(2006)]{2006ApJ...641..878T} Tasker, E.~J., \& Bryan, G.~L.\ 2006, \apj, 641, 878 

\bibitem[Toniazzo \& Schindler(2001)]{2001MNRAS.325..509T} Toniazzo, T., \& Schindler, S.\ 2001, \mnras, 325, 509 


\bibitem[Tonnesen \& Bryan(2012)]{2012MNRAS.422.1609T} Tonnesen, S., \& Bryan, G.~L.\ 2012, \mnras, 422, 1609 


\bibitem[Tonnesen \& Bryan(2010)]{2010ApJ...709.1203T} Tonnesen, S., \& Bryan, G.~L.\ 2010, \apj, 709, 1203-1218 


\bibitem[Tonnesen \& Bryan(2009)]{2009ApJ...694..789T} Tonnesen, S., \& Bryan, G.~L.\ 2009, \apj, 694, 789 


\bibitem[Tonnesen et al.(2011)]{2011ApJ...731...98T} Tonnesen, S., Bryan, G.~L., \& Chen, R.\ 2011, \apj, 731, 98 


\bibitem[Tonnesen \& Stone(2014)]{2014ApJ...795..148T} Tonnesen, S., \& Stone, J.\ 2014, \apj, 795, 148 


\bibitem[Turk et al.(2011)]{2011ApJS..192....9T} Turk, M.~J., Smith, B.~D., Oishi, J.~S., et al.\ 2011, \apjs, 192, 9 


\bibitem[van den Bosch et al.(2008)]{2008MNRAS.387...79V} van den Bosch, F.~C., Aquino, D., Yang, X., et al.\ 2008, \mnras, 387, 79 


\bibitem[Vollmer(2009)]{2009A&A...502..427V} Vollmer, B.\ 2009, \aap, 502, 427 


\bibitem[Vollmer(2003)]{2003A&A...398..525V} Vollmer, B.\ 2003, \aap, 398, 525 


\bibitem[Vollmer et al.(2001)]{2001ASPC..240..583V} Vollmer, B., Cayatte, V., Balkowski, C., \& Duschl, W.~J.\ 2001, Gas and Galaxy Evolution, 240, 583 


\bibitem[Vollmer et al.(2012)]{2012A&A...537A.143V} Vollmer, B., Soida, M., Braine, J., et al.\ 2012, \aap, 537, A143 


\bibitem[Vollmer et al.(2008)]{2008A&A...483...89V} Vollmer, B., Soida, M., Chung, A., et al.\ 2008, \aap, 483, 89 


\bibitem[Vollmer et al.(2006)]{2006A&A...453..883V} Vollmer, B., Soida, M., Otmianowska-Mazur, K., et al.\ 2006, \aap, 453, 883 


\bibitem[Warmels(1988)]{1988A&AS...72..427W} Warmels, R.~H.\ 1988, \aaps, 72, 427 


\bibitem[Yagi et al.(2010)]{2010AJ....140.1814Y} Yagi, M., Yoshida, M., Komiyama, Y., et al.\ 2010, \aj, 140, 1814 


\bibitem[Yoshida et al.(2008)]{2008ApJ...688..918Y} Yoshida, M., Yagi, M., Komiyama, Y., et al.\ 2008, \apj, 688, 918-930 



\end{thebibliography}
\end{document}